\documentclass{epl}

\title{Multi-excitons in self-assembled InAs/GaAs quantum dots: A
pseudopotential, many-body approach}
\author{A.J.~Williamson\inst{*} \and A.Franceschetti \and A.Zunger}
\institute{National Renewable Energy Laboratory, Golden, Colorado 80401}

\begin{document}
\maketitle
\begin{abstract}
We use a many-body, atomistic empirical pseudopotential approach to
predict the multi-exciton emission spectrum of a lens shaped InAs/GaAs
self-assembled quantum dot.  We discuss the effects of (i) The direct
Coulomb energies, including the differences of electron and hole
wavefunctions, (ii) the exchange Coulomb energies and (iii)
correlation energies given by a configuration interaction calculation.
Emission from the groundstate of the $N$ exciton system to the $N-1$
exciton system involving $e_0\rightarrow h_0$ and $e_1\rightarrow h_1$
recombinations are discussed.  A comparison with a simpler
single-band, effective mass approach is presented.
\end{abstract}

\vskip -1cm 
High-resolution single-dot
spectroscopy\cite{dekel98,dekel2000,landin99,toda99,zrenner2000} of
InAs/GaAs self-assembled quantum dots shows that as the excitation
intensity is increased, thus loading more excitons into the dots, new
emission lines appear both to the red and to the blue of the
fundamental emission line observed at low excitation power.  ``State
filling'' effects, leading to the recombination of high energy
electron-hole pairs, cannot explain the red-shifted emission lines,
nor the fact that the number of lines exceeds the number of allowed
single-particle transitions.  In this letter we present a theory of
self-assembled semiconductor quantum-dots, based on a pseudopotential
many-body expansion that demonstrates that it is multi-exciton
transitions that are responsible for this complex observed spectral
structure.  We isolate and clarify three distinct physical effects;
(i) electron-hole wavefunction asymmetry, leading to a {\em blue
shift} of the fundamental exciton transition as the number of
spectator excitons loaded into the dot increases, (ii)
electron-electron and hole-hole exchange interactions which {\em red
shift} all {\em even} multiexciton decays (biexciton, four-exciton)
and split the {\em odd} multiexciton decays (tri-exciton,
five-exciton) into multiple-lines and (iii) correlation effects which
{\em red shift} the biexciton leading to its binding with respect to
the monoexciton.

\begin{figure}[hbt]
\onefigure[height=6cm,width=13cm]{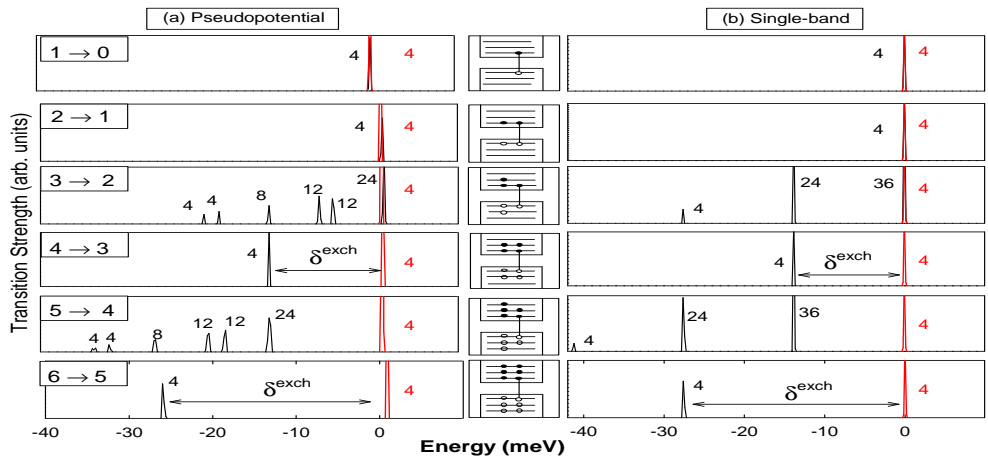}
\caption{Energy of $e_0\rightarrow h_0$, recombinations in the
presence of 0 to 5 spectator excitons.  Energies calculated in the
($\epsilon+J$)(red) and ($\epsilon+J+K$) (black) approximations are
shown.  (a) Shows our pseudopotential calculations and (b) shows
calculations with the assumptions from Refs.
\protect\cite{dekel98,barenco95,hawrylak99}.  The multiplicities of
each line is labelled.}
\label{barenco-comp}
\end{figure}

We will first describe the qualitative picture of multi-excitons and
then describe a quantitative model.  The essential physics of such
transitions can be understood by considering what happens to the
ground-state recombination of the lowest electron level, $e_0$, and
the lowest hole level, $h_0$, if other electrons and holes are present
in the dot as ``spectators''.  The schematic figures in the center of
Fig.~\ref{barenco-comp} depict the fundamental $e_0-h_0$ recombination
in the presence of 0 to 5 ``spectator'' electron-hole pairs (we assume
here that all levels are spatially
non-degenerate\cite{williamson2000:2}).  We distinguish here two
exciton series; (i) when the initial number of excitons, $N$ is even,
the initial electron configuration is ``closed shell'',
e.g. $(e_0^2)(h_0^2)$ for $N=2$, whereas (ii) when $N$ is odd, the
initial configuration has an open shell both in the electron and in
the hole manifold, e.g. $(e_0^2)e_1^1(h_0^2)h_1^1$ for $N=3$,
(parentheses are used for the closed shell orbitals).  The distinction
between the ``closed shell'' and ``open shell'' multiexciton is
important, since in the initial state of the $N$=even, ``closed
shell'' series the spectator levels can be occupied in only a single
unique manner by the spectator spins, while in the $N$=odd,
``open-shell'' series, many spin arrangements are possible in the
initial and final states.  This will lead to a large number of exciton
lines.

To understand qualitatively the effects of spectator electrons and
holes, let us consider the Hartree-Fock (HF) energy of a single
configuration of electrons and holes.  We denote the direct Coulomb
interaction between carriers in levels $i$ and $j$ by $J_{ij}$ and the
exchange interactions as $K_{ij}$.  The recombination energy of the
mono-exciton is given by
\begin{equation}\label{mono}
E^{1\rightarrow
0}_{e_0h_0}=(\epsilon_{e_0}-\epsilon_{h_0})-J_{e_0,h_0} \;\;\;,
\end{equation}
where $\epsilon_{e_0}$ and $\epsilon_{h_0}$ are the single particle
levels and $E^{1\rightarrow 0}_{e_0h_0}$ denotes the energy associated
with decaying from a single exciton to the groundstate.  Neglecting
the electron-hole exchange (NB. it is included in the calculation) the
$e_0-h_0$ recombination energy in the presence of $N_s$ electrons and
holes is
\begin{eqnarray}\label{n_to_n-1}
E^{N\rightarrow N-1}_{e_0h_0} & = & E^{1/0}_{e_0h_0}+ \left[
\sum_{e_s}^{N_s} \left(J_{e_0e_s}-J_{e_sh_0}\right) \right. \\ & + &
\left. \sum_{h_s}^{N_s} \left(J_{h_0h_s}-J_{e_0h_s}\right) \right] -
\left[ \sum_{e_s}^{N_s} K_{e_oe_s} +\sum_{h_s}^{N_s} K_{h_oh_s}
\right] \nonumber \;\;\;,
\end{eqnarray}
where $e_s$ and $h_s$ are spectator electrons and holes, such that
$\sum_{e_s}^{N_s}=\sum_{h_s}^{N_s}=N-1$.  We see from
Eq.(\ref{n_to_n-1}) that the $e_0-h_0$ recombination energy is shifted
with respect to the fundamental exciton.  This shift has two sources
indicated by the two bracketed terms in Eq.(\ref{n_to_n-1}).  First,
the Coulomb shift, which vanishes if the electrons and holes, $e_0$
and $h_0$ have the same wavefunctions, i.e. if $J_{e_0e_s}=J_{e_sh_0}$
and $J_{h_0h_s}=J_{e_0h_s}$.  This ``Coulomb shift'',
$\delta_{N\rightarrow N-1}^{Coul}$ thus reflects the difference in the
electron and hole wavefunctions.  It vanishes artificially in
single-band effective-mass calculations that use an infinite well
depth, such as those in Refs.\cite{dekel98,dekel2000,barenco95}.
Second, there is an exchange shift, $\delta_{N\rightarrow
N-1}^{exch}$, given by the second term in brackets in
Eq.(\ref{n_to_n-1}).  This exchange shift is familiar from theories of
band gap renormalization\cite{ambigapathy97} where the existence of
high carrier densities during high power photoexcitation act to reduce
the band gap.  In addition, since the exchange interaction depends on
the spin orientation of the carriers, the exchange contribution
[second term in Eq.(\ref{n_to_n-1})] can split the excitonic
transitions.

In the $N$=even, ``closed shell'' series, the initial state contains
no open shells while the final states contains one open shell.  This
results in one- and four-fold degeneracies for the initial and final
states, and hence 4 transitions.  These 4 transitions are split by the
small electron-hole exchange interaction.  In contrast, for the
$N$=odd transitions, the initial and final states contain one and two
open shells respectively, resulting in four- and 16-fold degeneracies
and hence a total 64 possible transitions.  Different alignments of
the spins produce splittings of the different transitions resulting
from electron-electron and hole-hole exchange interactions.  In
summary, within the HF approximation the presence of ``spectator
excitons'' will shift the $e_0-h_0$ transition in the $N$=even series
and shift and split the $e_0-h_0$ transition in the $N$=odd series.

The above treatment neglects the effects of correlation.  These can be
conveniently introduced by considering configuration-interaction (CI)
effects\cite{franceschetti99}.  In other words, instead of evaluating
the energy [Eq.(\ref{mono})] of the $e_0-h_0$ monoexciton by
considering only the 4 spin arrangements consistent with a {\em
single} orbital configuration, $e_0^1h_0^1$, we allow the presence of
other orbital configurations such as $e_1^1h_0^1$, $e_0^1h_1^1$ and
$e_2^1h_0^1$.  This configuration interaction approach can shift the
Hartree-Fock transitions of Eqs.(\ref{mono}) by a ``correlation
shift'', $\delta_{e_0h_0}^{CI}$, which includes both self-consistent
adjustments of the single particle orbitals and correlation.  This
will, for example, provide additional binding to the biexciton with
respect to the monoexciton
\begin{equation}\label{biexciton}
E^{2\rightarrow 1}_{e_0h_0}-E^{1\rightarrow
0}_{e_0h_0}=\left[J_{e_0,e_0}^{ee}+J_{h_0,h_0}^{hh}-2J_{e_0,h_0}^{eh}\right]
- \delta_{e_0h_0}^{CI} \;\;\;.
\end{equation}
Correlation effects can also add new transitions to the spectrum, due
to mixing of new configurations into the Hartree Fock configuration.
Thus, CI will, in general, also alter the intensity of the HF
transitions.

The purpose of this letter is to make a realistic prediction of the
``Coulomb shift'', $\delta^{Coul}$, the ``exchange-splitting',
$\delta^{Exch}$, ' and ``correlation shift/splitting'', $\delta^{CI}$,
due to the presence of spectator excitons in a self-assembled
semiconductor quantum dot.  We adopt the experimentally determined
lens shaped dot with a base of 250 \AA~and a height of 35 \AA and a PL
peak at 1.1 eV.  The optical and electronic properties of this dot
geometry have been extensively studied, see
Refs.\cite{drexler94,warburton98} and references therein.  Recently,
alternative dot geometries for {\em uncapped} dots have been proposed,
based on \{136\} facets\cite{lee98}.  However, there is strong
evidence of Ga in-diffusion during the capping process which acts to
produce a lens shaped geometry.  The single-particle bound states are
calculated using an empirical pseudopotential
Hamiltonian\cite{williamson2000:2}.  Recent developments in these
pseudopotentials enable the full inclusion of the effects of strain,
multi-band couplings, band non-parabolicity and spin-orbit coupling in
the single particle Schr\"{o}dinger equation.  In
Ref.\cite{williamson2000:2} we demonstrate that by accurately fitting
the {\em bulk} band structure, effective masses and deformation
potentials we are able to use these pseudopotentials to obtain
excellent agreement with a wide range of optical and electronic
properties for the lens shaped dots discussed in this paper.  These
comparisons were not possible earlier when the shape of the dot and
its composition profile were unknown.  In Ref.\cite{williamson2000:2},
we model the shape and composition profile of the dot and then compare
calculated and measured single exciton energies, inter-band energy
spacings, electron and hole binding energies and wetting layer
energies, finding excellent agreement.  For completeness, we treat
here the same lens-shape, alloyed dot with a 1.1 eV PL peak.
Unfortunately, although numerous experiments were conducted on this
dot, no multi-exciton spectra were taken since the lowest PL is
outside the range of conventional CCD detection equipment, so
quantitative comparison awaits a future measurement.

Having obtained the single-particle levels ($e_0,e_1$...,
$h_0,h_1$...) from the pseudopotential method, we then calculate
numerically the screened Coulomb and exchange integrals
\begin{eqnarray}\label{Jeh}
J_{ijkl} & = & \int\int\frac{\psi_i^*({\bf r}_1)\;\psi_j({\bf
r}_2)\;\psi_k^*({\bf r}_1)\;\psi_l({\bf
r}_2)}{\overline{\epsilon}({\bf r}_1-{\bf r}_2)|{\bf r}_1-{\bf r}_2|}
\;d{\bf r}_1 d{\bf r}_2 \;\;\;, \nonumber \\ K_{ijkl} & = &
\int\int\frac{\psi_i^*({\bf r}_1)\;\psi_j({\bf r}_2)\;\psi_k^*({\bf
r}_2)\;\psi_l({\bf r}_1)}{\overline{\epsilon}({\bf r}_1-{\bf
r}_2)|{\bf r}_1-{\bf r}_2|} \;d{\bf r}_1 d{\bf r}_2 \;\;\;,
\end{eqnarray}
where $\overline{\epsilon}$ is a size dependent, phenomenological,
screened dielectric function\cite{williamson2000:1}.  Our exchange
automatically includes both short and long range
components\cite{franceschetti98}.  In the configuration-interaction
approach, we expand the wavefunction of the $N$-exciton, $\Psi$, as a
linear combination of Slater determinants, $\Phi$, obtained by
exciting $N$ electrons from the valence band to the conduction band.
For example, in the $N=2$ (biexciton) case we have:
\begin{equation}
\Psi({\bf r}_1,\cdots,{\bf r}_M)=\sum_{vv',cc'} A_{vv',cc'} \Phi_{vv',cc'}({\bf r}_1,\cdots,{\bf r}_M) \;\;\;,
\end{equation}
Here $M$ is the total number of electrons in the quantum dot, $v,v'$
denote the unoccupied valence band single particle states (holes) and
$cc'$ denote the conduction band states occupied by the $N$ excited
electrons.  To obtain our many-body states, we diagonalize the
many-body Hamiltonian in the basis of $\Phi$.

To clarify and isolate the physical factors contributing to
multi-exciton effects we solve the problem in a series of steps.
First, we neglect configuration interaction effects, treating only
single-configurations and set all the exchange integrals to zero,
($\delta^{exch}=0$).  In this ($\epsilon+J$) approximation we see the
effect of the Coulomb ``chemical shift'', $\delta^{Coul}$.  Second, we
will still use a single-configuration, but include exchange integrals.
In this ($\epsilon+J+K$) approximation we will see the added effects
of carrier-carrier exchange, $\delta^{exch}$.  Finally, the effects of
correlation are included in the CI calculation.

Figure~\ref{barenco-comp}(a) shows  our calculated energies associated
with  the fundamental  recombination  of an  electron  and hole,  $e_0
\rightarrow h_0$ in the presence  of 0 to 5 spectator excitons.  These
are denoted  as the $1\rightarrow 0$ to  $6\rightarrow 5$ transitions.
The $N$  excitons that form the  initial state of  each transition are
assumed to  occupy the groundstate  configuration as predicted  by the
Aufbau principle.   A schematic energy  level scheme in the  center of
each figure,  shows this initial occupation, where  the vertical solid
line marks the recombination taking place.  The spectrum shown in each
panel is obtained from a sum  of Gaussians (0.1 meV width) whose means
represent the transition energies  and heights are proportional to the
calculated dipole transition  element.  The zero of energy  is taken as
the  fundamental  exciton, $\epsilon_{e_0}-\epsilon_{h_0}-J_{e_0h_0}$.
The  multiplicity  of each  individual  group  of  transitions in  the
($\epsilon+J$) and ($\epsilon+J+K$) approximations are marked in black
and red.

{\em The effect of direct Coulomb interactions: } The red lines in
Fig.~\ref{barenco-comp}(a) show the transition energies in the
($\epsilon+J$) approximation which include only single particle and
direct Coulomb energies.  Inspection of these lines shows that:

(i) All the observed Coulomb shifts are relatively small,
($\delta^{Coul}\sim 2$ meV) and result in a blue shift of the
transition with respect to the fundamental transition.  For example,
at this level of approximation, the biexciton is ``unbound'' with
respect to two single excitons, i,e. $E^{2\rightarrow
1}>E^{1\rightarrow 0}$.

(ii) As all the transitions involve the same single particle $e_0
\rightarrow h_0$ recombination, they all have the same oscillator
strength.

(iii) All the transitions show only a single degenerate line as
neither the initial or the final state exhibit any exchange splittings
in this approximation.

{\em The effect of exchange interactions: } The black lines in
Fig.~\ref{barenco-comp}(a) show the transition energies in the
($\epsilon+J+K$) approximation which includes single particle, direct
and exchange Coulomb energies.  Inspection of these lines shows that:

(i) The $1\rightarrow 0$ and $2\rightarrow 1$ transitions contain only
one or zero spectator excitons and hence no electron-electron or
hole-hole exchange takes place.  Therefore, only the small
electron-hole exchange fine splitting is observed.  We therefore
classify these transitions as essentially 4-fold multiplets.  Due to
the lack of correlation the biexciton is still unbound with respect to
two single excitons as in the ($\epsilon+J$) approximation.

(ii) For $even\rightarrow odd$ there are 4 possible transitions.  In
each case, $2\rightarrow 1$, $4\rightarrow 3$ and $6\rightarrow 5$ we
observe only one 4-fold multiplet.  The fine splitting within this
group of transitions arises from electron-hole exchange splittings of
the 4 final states.  The initial states contain no unpaired electrons
and holes and hence exhibit no exchange splittings.  As the number of
spectator excitons increases, there is a red shift of these
transitions due to exchange interactions.  The exchange energy shifts
from Eq.(\ref{n_to_n-1}) are $\delta^{exch}_{4\rightarrow
3}=[K_{e_0e_1}+K_{h_0h_1}]$ and $\delta^{exch}_{6\rightarrow
5}=[K_{e_0e_2}+K_{e_0e_1}+K_{h_0h_2}+K_{h_0h_1}]$.

(iii) For $odd\rightarrow even$ with $N\geq 3$ there are 64 possible
transitions.  In both cases: $3\rightarrow 2$ and $5\rightarrow 4$ we
see 6 groups of transitions with multiplicities of 4:4:8:12:12:24.
The splitting between the 6 groups arises from electron-electron and
hole-hole exchange splittings between the final states which contain
two unpaired electrons and holes.  The 6 groups of transitions in
$3\rightarrow 2$ and $5\rightarrow 4$ span 23 and 22 meV.  This energy
span reflects the span of the 16 eigenvalues of the 16x16 matrix
generated by the different spin occupations of the final states.

(iv) The exchange interaction alters the oscillator strength of the
transitions so that they are not all identical as in the
($\epsilon+J$) approximation.

{\em Comparison with previous calculations:} Previous calculations of
multi-excitons in quantum dots include the works of
Hu\cite{hu90},Takagahara\cite{takagahara99}, Barenco and
Dupertuis\cite{barenco95}, Dekel {\em
et. al.}\cite{dekel98,dekel2000}, Landin{\em et. al.}\cite{landin99}
and Hawrylak\cite{hawrylak99}.
Refs.\cite{hu90,barenco95,dekel98,dekel2000,takagahara99} and
\cite{hawrylak99} adopt single-band effective-mass models with either
an infinite potential
barrier\cite{hu90,barenco95,dekel98,dekel2000,takagahara99} or a
parabolic potential\cite{hawrylak99}, both of which artificially force
the electron and hole wavefunctions to be identical.  Therefore, in
all these calculations the Coulomb shift, $\delta^{Coul}$, is zero and
the electron-hole exchange vanishes, $K_{e_ih_j}=0$.  To obtain
realistic single particle energy spacings in
Refs.\cite{barenco95,dekel98,dekel2000} an unrealistic cuboidal shape
had to be assumed and the size of the dots was treated as adjustable
parameters.  By choosing different lengths for all three dimensions
both the measured $s$-$p$ and $p$-$p$ splittings can be reproduced.
The two-dimensional parabolic potential adopted in
Ref.\cite{hawrylak99} can also be adjusted to reproduce the correct
$s$-$p$ splitting, but will always produce degenerate $p$ states.  It
has recently been shown\cite{dekel2000} that only models which can
split this $p$ level degeneracy can provide a realistic interpretation
of experimental results.  In addition to approximating the single
particle states,
Refs.\cite{hu90,barenco95,dekel98,dekel2000,takagahara99,hawrylak99}
neglect the effects of strain and the spin-orbit interaction.

To calculate the exchange and correlation contribution to the
excitonic energies Ref.\cite{hu90} uses path integral quantum Monte
Carlo techniques which provide an exact (to within statistical error)
determination of the exchange and correlation energy.  However,
quantum Monte Carlo methods are currently restricted to single band
Hamiltonians and cannot therefore predict the Coulomb shift,
$\delta^{Coul}$, discussed above.  Ref.\cite{hawrylak99} adopts a
limited basis CI to estimate correlation energies.
Refs.\cite{barenco95,dekel98,dekel2000,takagahara99,landin99} use only
a single configuration approach which does not include correlation
effects.

To assess the above approximations used in
\cite{hu90,barenco95,dekel98,dekel2000,hawrylak99} we show in
Fig.~\ref{barenco-comp}(b) a repeat of our calculations for the
transition energies within the ($\epsilon+J$) and ($\epsilon+J+K$)
approximations applying the assumptions adopted in
Refs.\cite{dekel98,dekel2000,barenco95,hawrylak99}, namely
$J_{e_ie_j}=J_{h_ih_j}=J_{e_ih_j}, K_{e_ie_j}=K_{h_ih_j},
K_{e_ih_j}=0$ and $\Delta^{SO}=0$.  We observe that:

(i) Within the ($\epsilon+J$) approximation, the chemical shift,
$\delta^{Coul}$ is zero by definition so all the red lines lie on the
zero of energy.  (ii) Within the ($\epsilon+J+K$) approximation for
$even\rightarrow odd$, there are 4 possible transitions that are
exactly degenerate as $K_{e_ih_j}=0$.  (iii) For $odd\rightarrow
even$, there are 64 possible transitions, split into 4:24:36
multiplets, compared to the 4:4:8:12:12:24 multiplets obtained in the
pseudopotential calculations.  The reduction in the number of
multiplets arises form the assumptions $\psi_{e_i}=\psi_{h_i}$ and
$\Delta^{SO}=0$.  Other than these changes we find that many of the
qualitative feature noted in the calculations of Dekel {\em
et. al.}\cite{dekel98,dekel2000} are retained in the pseudopotential
description.

\begin{figure}[hbt]
\oneimage[height=6.5cm,width=13cm]{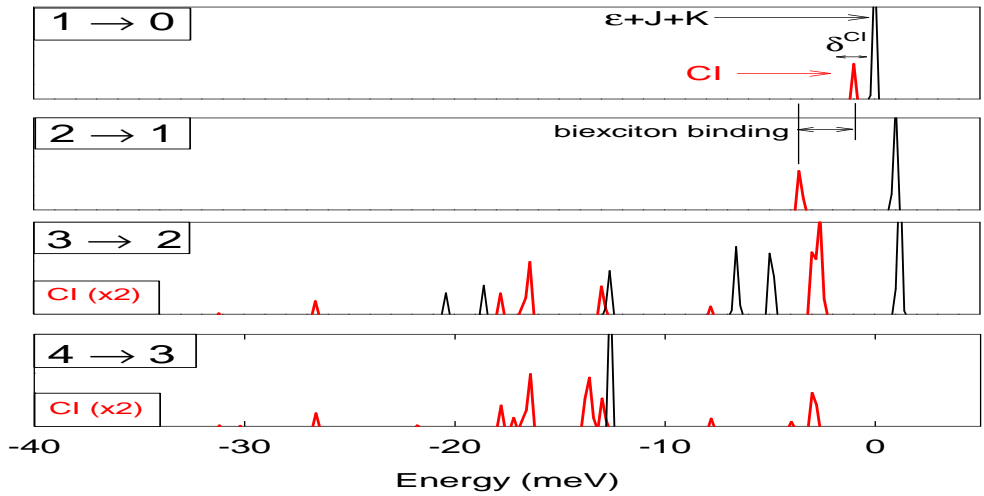}
\caption{Energy of $e_0\rightarrow h_0$, recombinations in the
presence of 0 to 3 spectator excitons.  The black and red lines show
energies calculated in the ($\epsilon+J+K$) and CI approximations.}
\label{A-manifold}
\end{figure}

{\em The effect of CI interactions: } In Fig.~\ref{A-manifold} we
contrast the transition energies from our pseudopotential calculations
within the $(\epsilon+J+K)$ approximation (black lines) with those
from a CI calculation (red lines) which includes single particle,
direct and exchange Coulomb and ``correlation'' effects.  

In the CI calculations we expand the many-body wavefunction in a basis
of Slater determinants constructed from all possible orbital and spin
occupations of the lowest 10 (including spin degeneracy) electron and
lowest 10 hole single particle levels.  For example, the biexciton
basis contains $^{10}C_{2}.^{10}C_{2}=2025$ Slater determinants, and
the 3-exciton basis contains $^{10}C_{3}.^{10}C_{3}=14400$ Slater
determinants.  This basis neglects the contributions for higher lying
bound states and continuum states.  To investigate the effects of
adopting this limited CI basis\cite{shumway2000:1} we have compared
the results of diffusion quantum Monte Carlo calculations (DMC) and CI
calculations for the exciton correlation energy in a model, single
band system with equivalent size, band offsets and number of bound
states.  We find that our CI calculations retreive approximately
50-60\% of the 5-6 meV of correlation energy obtained in the DMC
calculations and can therefore be used as a {\em lower bound} for the
effects of correlation.  Using these calculations we find that

(i) Correlation effects lower the energy of both the initial and final
state of a transition.  Our calculated correlation varies from 2-3 meV
for a single exciton to 10 meV for multiple excitons.  For all the
transitions shown here, the correlation shift, $\delta^{CI}$, for the
initial state with $N$ excitons is greater than that for the final
$N-1$ exciton state.  Therefore, all the CI {\em transition} peaks are
red shifted with respect to those from the ($\epsilon+J+K$)
approximation.  (ii) This ($\delta^{CI}$) red shift is larger for the
$2\rightarrow 1$ transition than for the $1\rightarrow 0$ transition
and is able to overcome the Coulomb blue shift and ``bind'' the
biexciton.  (iii) As the number of spectator excitons increases, the
{\em difference} in the red shift for the initial and final states
decreases, so that the red shift of the {\em transition} energy
decreases.  For $3\rightarrow 2$ and $4\rightarrow 3$ transitions the
red shifts of the transitions rapidly decrease.  (iv) The mixing of
configurations within the CI results in both additional peaks in the
CI spectra and changes in the relative magnitude of the peaks,
e.g. for $4\rightarrow 3$ the single ($\epsilon+J+K$) peak is split
into two equally strong multiplets of peaks in the CI spectra.  The
additional peaks result from the mixing in of excited states within
the CI calculation.  For example, the additional CI peaks blue shifted
from the main peaks in the $4\rightarrow 3$ spectra result from
excited states configurations mixed into the $N=3$ exciton.

In conclusion, we present results of the first pseudopotential,
many-body calculation of multi-exciton states within InAs/GaAs
quantum dots.  We are able to isolate the effects of the direct and
exchange Coulomb interactions and correlation on the energies of
$N\rightarrow N-1$ excitonic transitions.  We find that direct
Coulomb energies introduce small blue shifts.  Electron-electron and
hole-hole exchange splittings which are responsible for the majority
of the observed structure, introduce both red shifts and splittings.
Correlation effects red shift all transitions and change the relative
energies of transitions (e.g. bind the biexciton).

We thank E. Dekel and D. Gershoni for many
useful discussions.  This work was supported DOE -- Basic Energy
Sciences, Division of Materials Science under contract
No. DE-AC36-99GO10337.  AZ acknowledges support from the Binational
US-Israel Science Foundation (453/97).


$^*$Present Address: Lawrence Livermore National Laboratory, CA 94550

\vskip 0.5 cm

\end{document}